\documentclass[a4paper,11pt]{article}
\usepackage[english]{babel}
\usepackage{amsmath,amsthm}
\usepackage{amsfonts}
\usepackage{graphicx}
\usepackage[top=1in,bottom=1in,left=1.25in,right=1in]{geometry}

\theoremstyle{definition}

\theoremstyle{remark}

\numberwithin{equation}{section}
\begin{document}

\title{Symmetric Ternary Quantum Homomorphic Encryption Schemes Based on the Ternary Quantum One-Time Pad}
\author{Yuqi Wang,Kun She,Qingbin Luo,Fan Yang, and Chao Zhao}

\maketitle
\begin{abstract}
  Aiming at a ternary quantum logic circuit, four symmetric ternary quantum homomorphic encryption schemes, based on ternary quantum one-time protocol, were presented. First, for a one-qutrit rotation gate, a homomorphic quantum encryption scheme was constructed. Second, in view of the synthesis of a general $3 \times 3$ unitary transformation, another one-qutrit quantum homomorphic encryption scheme was proposed. Third, according to the one-qutrit scheme, the two-qutrit quantum homomorphic encryption scheme about GCX($\emph{m'})$gate was constructed and was further generalized to the \emph{n}-qutrit unitary matrix case. Finally, the security of these schemes was analyzed from two perspectives. It could be concluded that the attacker can correctly guess the encryption key with a maximum probability ${p_k} = {1 \mathord{\left/
 {\vphantom {1 {{3^{3n}}}}} \right.
 \kern-\nulldelimiterspace} {{3^{3n}}}}$, thus it can better protect the privacy of users’ data. Moreover these schemes can be well integrated into future quantum remote server architecture, and the computational security of the user’s private quantum information can be well solved in a distributed computing environment.
\end{abstract}

\section{Introduction}

In a distributed computing environment, customers have a large amount of data stored in the remote server. These data may include personal bank account information, online shopping records, credit card consumption records, etc., and this information belongs to customers’ private encrypted data which is indistinguishable for a remote server. Suppose we intend to compute on the encrypted data without a decryption process, or delegate it to a trusted third party without leaking information of the input data: is it possible to do so? Fortunately, blind computation\cite{G10}  or homomorphic encryption\cite{G09,BV14,BGV14,DGHV10} can achieve it perfectly without a decryption process and leaking private information of the encrypted input data.

From the perspective of quantum information processing, performing operations on encrypted data without a decryption process is relative to blind quantum computation\cite{BFKW13,GMMR13,MPF13,FK14,MF13,FBSYLPJR14}and quantum homomorphic encryption (QHE). This paper will study this problem by the QHE technique that can not only protect the privacy of users’ data, but also accomplish secure computatiom on a remote server.

Rohde et al.\cite{RFG12} firstly studied quantum walks with encrypted data, and then they proposed a limited QHE scheme using the Boson sampling and multi-walker quantum walk models on Linear Optics Quantum Computation. However, QHE has still not been defined, and a quantum fully homomorphic encryption (QFHE) scheme has not yet been constructed.

Liang\cite{L13}  firstly presented the definitions of the QHE and QFHE, and then, based on the Quantum One-Time Pad (QOTP) protocol, he constructed the symmetric QHE and QFHE scheme with perfect security, where evaluation function depends on the encryption keys. Subsequently, learning from the Universal Quantum Circuit (UQC), he proposed a QFHE scheme\cite{L14}. In the scheme, the encryption key is different from the decryption key, and cannot be public. Moreover, evaluation algorithm is independent of the encryption key, and the decryption key can be computed from the encryption key by an interactive update process. Recently, Liang\cite{LY15} again presented two QFHE schemes, which are constructed based on quantum fault-tolerant construction. The characteristics are using quantum CSS code as the secret key and containing the periodical interaction between Client and Server.

Armknecht et al.\cite{AGKP14}  proved the general impossibility of (Abelian) group homomorphic encryption in the presence of quantum adversaries, when assuming the IND-CPA security notion as the minimal security requirement. And they provided a sufficient condition and discussed its satisfiability in non-group homomorphic cases. Tan et al.\cite{TKOCF15}  presented a private-key QHE scheme that hides arbitrary quantum computations. A particular instance of their encoding hides information at least proportional to $m\log m$ bits when m bits are encrypted. Recently, Broadbent et al.\cite{BJ15} presented QHE schemes for circuits of low T-gate complexity. These schemes allow for arbitrary Clifford group gates, but they become inefficient for circuits with a large degree of complexity, measured in terms of the non-Clifford portion of the circuit.

Currently, QHE research is limited and mainly focused on the quantum bits (qubits). According to present research, this article presented four Ternary QHE (TQHE) schemes based on the Ternary QOTP (TQOTP) for the first time. The rest of the paper is organized as follows. Section 2 provides a brief introduction to ternary quantum gates and QHE, and a TQOTP scheme is presented. In section 3, the first TQHE scheme, based on one-qutrit rotation gates$R_\partial ^{(ij)}(\theta )$, is constructed, and it is generalized to a general ternary quantum gate with synthesis idea. Then the third TQHE scheme about a GCX gate is constructed, and extended to \emph{n}-qutrit gate in theory. These schemes are analyzed by synthesizing concepts from ternary quantum gates, secret key security and user data privacy in section 4. Conclusions and future research ideas are presented in section 5.

\section{Preliminaries}

\subsection{Universal Quantum Circuits}

\textbf{Definition 1.} (Universal Quantum Circuit (UQC)\cite{BFGH10}) Fix $n>0$ and let \emph{C} be a collection of quantum circuits on \emph{n} qubits. A quantum circuit \emph{U} on $(n+m)$ qubits is universal for \emph{C} if, for every circuit ${C_U} \in C$, there is a string $x\in{\{0,1\}^m}$(the encoding) such that, for all strings,$y \in {\{ 0,1\} ^n}$   (the data),
	
\begin{equation}\label{e1}
{U(\left| y \right\rangle  \otimes \left| x \right\rangle ) = {C_U}(\left| y \right\rangle ) \otimes \left| x \right\rangle }
\end{equation}

The definition of the UQC tells us two things: one is that an arbitrary \emph{U} transformation can be synthesized by a finite number of logic gates of the \emph{C} set. Another is that blind quantum computation or homomorphic encryption scheme can be constructed, which will be discussed later. In the UQC, given a \emph{n}-qubit $\left| y \right\rangle$ as the input data, while a \emph{m}-qubit $\left| x \right\rangle$  is input as the encoding of a quantum transformation ${C_U} \in C$, the UQC would output $(n+m)$ qubits (shown in Eq.\ref{e1}). Here,\emph{n}-qubit $\left| x \right\rangle$ is called the encoding of the quantum transformation ${C_U} \in C$ with regard to the UQC \emph{C}. $\left| x \right\rangle$ as input data can not only hide the data $\left| y \right\rangle $ but also protect the operation $ {C_U}$. Furthermore, Eq.\ref{e1} gives a transformation between \emph{U} and ${C_U}$ quantum circuit, which means it is possible to construct a homomorphism operator between \emph{U} and ${C_U}$.

\subsection{Ternary Quantum Circuit}

A qutrit is represented as a unit vector in state space, which is a complex three-dimensional vector space ${\emph{H}_3} $. In the computational basis, the basis vectors (or basis states) of ${\emph{H}_3} $ are written in Dirac notation as $\left| 0 \right\rangle ,\left| 1 \right\rangle$, and $\left| 2 \right\rangle $, where $\left| 0 \right\rangle  \equiv {(1,0,0)^T}$, $\left|1\right\rangle  \equiv {(0,1,0)^T}$ , and $\left| 2 \right\rangle  \equiv {(0,0,1)^T}$ . An arbitrary vector $\left| \varphi  \right\rangle $ in ${\emph{H}_3}$ can be expressed as a linear combination $\left| \varphi  \right\rangle  = {a_0}\left| 0 \right\rangle  + {a_1}\left| 1 \right\rangle  + {a_2}\left| 2 \right\rangle $, where ${a_i} \in \textbf{C}$ and $\sum\nolimits_{i = 0}^2 {|{a_i}{|^2}}  = 1$. The real number $|{a_i}{|^2}$ is the probability that the state vector   will be in the \emph{i}th basis state upon measurement.

A qudit is represented as a unit vector in the state space, which is a complex projective \emph{d} dimensional Hilbert space ${\emph{H}_d}$. In the computational basis, the basis vectors of ${\emph{H}_d}$ are written in Dirac notation as $\left| 0 \right\rangle ,\left| 1 \right\rangle , \cdot  \cdot  \cdot ,\left| {d - 1} \right\rangle $, where $\left| i \right\rangle $ =${(0, 0, . . . , 1, . . . , 0)^T} $ with a 1 in the (\emph{i}+1)st coordinate, for $0 \le i \le d - 1$. Its linear combination is similar with that of the qutrit. Note that the basis vectors in the computational basis are ordered by natural numbers.

\subsubsection{Common one-qutrit circuits}

1)Ternary ${X^{(ij)}}$  (TX) gates

The operators ${X^{(ij)}}$ are defined as follows,
\begin{equation}\label{e2}
 {X^{(ij)}} = \left| i \right\rangle \left\langle j \right| + \left| j \right\rangle \left\langle i \right| + \sum\nolimits_{k \ne i,j} {\left| k \right\rangle \left\langle k \right|,i,j \in \{ 0,1,2\} }
\end{equation}
The ${X^{(ij)}}$ gates acting on $\left|k\right\rangle $ can be represented as ${X^{(ij)}}\left| k \right\rangle  = \left\{ {\begin{array}{*{20}{c}}
{\left| j \right\rangle ,\;\;k = i}\\
{\left| i \right\rangle ,\;\;k = j}\\
{\left| k \right\rangle ,others}
\end{array}} \right.$. Because ${X^{(ij)}} = {X^{(ji)}}$  and ${X^{(ij)}}_{i = j} = {I_3}$,TX gates have three valid forms and are defined as follows:
\begin{equation}\label{e3}
 {X^{(0)}} \equiv {X^{(01)}},{X^{(1)}} \equiv {X^{(02)}},\;and\;{X^{(2)}} \equiv {X^{(12)}}
\end{equation}
2)Ternary ${H^{(ij)}}$ gates

The ternary extension of Hadamard gate can be expressed as
\begin{equation}\label{e4}
\begin{array}{l}
{H^{(0)}} \equiv {H^{(01)}} = {1 \mathord{\left/
 {\vphantom {1 {\sqrt 2 }}} \right.
 \kern-\nulldelimiterspace} {\sqrt 2 }}\left( {\begin{array}{*{20}{c}}
1&1&0\\
1&{ - 1}&0\\
0&0&{\sqrt 2 }
\end{array}} \right),\;{H^{(1)}} \equiv {H^{(02)}} = {1 \mathord{\left/
 {\vphantom {1 {\sqrt 2 }}} \right.
 \kern-\nulldelimiterspace} {\sqrt 2 }}\left( {\begin{array}{*{20}{c}}
1&0&1\\
0&{\sqrt 2 }&0\\
1&0&{ - 1}
\end{array}} \right),\\
{H^{(2)}} \equiv {H^{(12)}} = {1 \mathord{\left/
 {\vphantom {1 {\sqrt 2 }}} \right.
 \kern-\nulldelimiterspace} {\sqrt 2 }}\left( {\begin{array}{*{20}{c}}
{\sqrt 2 }&0&0\\
0&1&1\\
0&1&{ - 1}
\end{array}} \right).
\end{array}
\end{equation}
The ${H^{(ij)}}$ gates acting on $\left| k \right\rangle $  can be expressed as
\begin{equation}
{H^{(ij)}}\left| k \right\rangle  = \left\{ {\begin{array}{*{20}{c}}
{{1 \mathord{\left/
 {\vphantom {1 {\sqrt 2 }}} \right.
 \kern-\nulldelimiterspace} {\sqrt 2 }}(\left| i \right\rangle  + \left| j \right\rangle ),k = i}\\
{{1 \mathord{\left/
 {\vphantom {1 {\sqrt 2 }}} \right.
 \kern-\nulldelimiterspace} {\sqrt 2 }}(\left| i \right\rangle  - \left| j \right\rangle ),k = j}\\
{\left| k \right\rangle ,\quad \quad\quad\;\,\;k \ne i,j}
\end{array}} \right.
\end{equation}
3)Ternary ${Z^{(i)}}$ gates

The ${Z^{(i)}}$ gates can be denoted as follows,
\begin{equation}\label{e5}
{Z^{(0)}} \equiv diag\{  - 1,{I_2}\} ,{Z^{(1)}} \equiv diag\{ 1, - 1,1\} ,{Z^{(2)}} \equiv diag\{ {I_2}, - 1\}
\end{equation}
The $ {Z^{ij}}$ gates applied to $\left| k \right\rangle $ can be expressed as ${Z^{(i)}}\left| k \right\rangle  = \left\{ {\begin{array}{*{20}{c}}
{ - \left| k \right\rangle ,i = k}\\
{\left| k \right\rangle ,\;\;\,i \ne k}
\end{array}} \right.$.
\\
4)Ternary shift gates

The ternary shift gates are listed in Table 1, in which the addition is mode 3 addition.
\begin{table}[tbp]
\centering
\caption{\label{t1}{Ternary shift gates.}}
\begin{tabular}{lll}
\hline
Name &Operation &Equivalent to X operations\\ \hline
Buffer &$f(x)=x$ &${I_3}$\\
Single-shift &$f(x)=x+1$ &${X^{01}}{X^{12}}$ \\
Dual-shift &$f(x)=x+2$ &${X^{12}}{X^{01}}$ \\
Self-shift &$f(x)=2x$ &${X^{12}}$ \\
Self-Single-shift&$f(x)=2x+1$ &${X^{01}}$ \\
Self-Dual-shift &$f(x)=2x+2$ &${X^{02}}$ \\ \hline
\end{tabular}
\end{table}
\\
5)Ternary rotation gates

The definition of a one-qutrit rotation gate is listed in Eq.\ref{e6}.
\begin{equation}\label{e6}
R_\partial ^{(ij)}(\theta ) = \exp ( - i\sigma _\partial ^{(ij)}{\theta  \mathord{\left/
 {\vphantom {\theta  2}} \right.
 \kern-\nulldelimiterspace} 2})
\end{equation}
where $\partial  \in \{ x,y,z\} $, $0 \le i < j \le 2$, $\sigma _x^{(ij)} = \left| i \right\rangle \left\langle j \right| + \left| j \right\rangle \left\langle i \right|$, $\sigma _y^{(ij)} =  - i\left| i \right\rangle \left\langle j \right| + i\left| j \right\rangle \left\langle i \right|$, $\sigma _z^{(ij)} = \left| i \right\rangle \left\langle i \right| - \left| j \right\rangle \left\langle j \right|$. Based on exponent mapping\cite{DZW08}, the following four equations are obvious,$R_y^{(01)}(\theta ) = \left( {\begin{array}{*{20}{c}}
{\cos ({\theta  \mathord{\left/
 {\vphantom {\theta  2}} \right.
 \kern-\nulldelimiterspace} 2})}&{ - \sin ({\theta  \mathord{\left/
 {\vphantom {\theta  2}} \right.
 \kern-\nulldelimiterspace} 2})}&0\\
{\sin ({\theta  \mathord{\left/
 {\vphantom {\theta  2}} \right.
 \kern-\nulldelimiterspace} 2})}&{\cos ({\theta  \mathord{\left/
 {\vphantom {\theta  2}} \right.
 \kern-\nulldelimiterspace} 2})}&0\\
0&0&1
\end{array}} \right),R_y^{(02)}(\theta ) = \left( {\begin{array}{*{20}{c}}
{\cos ({\theta  \mathord{\left/
 {\vphantom {\theta  2}} \right.
 \kern-\nulldelimiterspace} 2})}&0&{ - \sin ({\theta  \mathord{\left/
 {\vphantom {\theta  2}} \right.
 \kern-\nulldelimiterspace} 2})}\\
0&1&0\\
{\sin ({\theta  \mathord{\left/
 {\vphantom {\theta  2}} \right.
 \kern-\nulldelimiterspace} 2})}&0&{\cos ({\theta  \mathord{\left/
 {\vphantom {\theta  2}} \right.
 \kern-\nulldelimiterspace} 2})}
\end{array}} \right)$, $R_z^{(01)}(\theta ) = \left( {\begin{array}{*{20}{c}}
{{e^{{{ - i\theta } \mathord{\left/
 {\vphantom {{ - i\theta } 2}} \right.
 \kern-\nulldelimiterspace} 2}}}}&0&0\\
0&{{e^{{{i\theta } \mathord{\left/
 {\vphantom {{i\theta } 2}} \right.
 \kern-\nulldelimiterspace} 2}}}}&0\\
0&0&1
\end{array}} \right)$  and $R_z^{(02)}(\theta ) = \left( {\begin{array}{*{20}{c}}
{{e^{{{i\theta } \mathord{\left/
 {\vphantom {{i\theta } 2}} \right.
 \kern-\nulldelimiterspace} 2}}}}&0&0\\
0&1&0\\
0&0&{{e^{{{ - i\theta } \mathord{\left/
 {\vphantom {{ - i\theta } 2}} \right.
 \kern-\nulldelimiterspace} 2}}}}
\end{array}} \right)$.

A one-qutrit gate is essentially a $3 \times 3$ unitary matrix. According to the Cartan Decomposition of Lie algebra, the unitary matrix U has the following form\cite{DW14},
\begin{equation}\label{e7}
\begin{array}{l}
\begin{array}{l}
{U_1} = {e^{i\alpha }}R_y^{(01)}(\beta )R_y^{(02)}(\gamma )R_y^{(01)}(\delta )R_z^{(01)}(\theta ),\\
{U_2} = R_z^{(02)}(\varphi )R_y^{(01)}(\beta ')R_y^{(02)}(\gamma ')R_y^{(01)}(\delta '),\\
U = {U_1} \times {U_2}.
\end{array}
\end{array}
\end{equation}
where $\alpha ,\beta ,\gamma ,\delta ,\theta ,\varphi ,\beta',\gamma'$ and $\delta'$ are all real numbers. The four basic one-qutrit rotation gates, $R_y^{(01)}(\beta ),R_y^{(02)}(\gamma ),R_z^{(01)}(\theta )$, and $R_z^{(02)}(\varphi )$, constitute a set of one-qutrit elementary gates\cite{DZW08}.

\subsubsection{Common two-qutrit circuits}

1)Ternary NOT gate

A Ternary NOT gate ${N_j}({A_i})$ is defined as ${P_j} = {A_i}{ \oplus _3}1,if\;j = i;else\;{P_j} = {A_i}$, where ${ \oplus _3}$ stands for addition modulo 3.
\\
2)Feynman gate

It is defined as $T{F_{A,B}}(A,B) = (A,A{ \oplus _3}B)$, where ${ \oplus _3}$  stands for addition modulo 3, and $A,B \in \left\{ {0,1,2} \right\}$.
\\
3)Ternary \emph{K}-Controlled \emph{X} (TKCX) Gate

For a $TKC{X_{{i_1},{i_2} \cdot  \cdot  \cdot ,{i_k};j}}(B)$ gate, when ${B_{i1}} = {B_{i2}} = \cdot  \cdot  \cdot = {B_{ik}} = 2,{X^{(ij)}}$ will be acted on target bit ${B_j}$, therein ${B_{i1}},{B_{i2}}, \cdot  \cdot  \cdot ,{B_{ik}}$ are control bits. When \emph{K}=1, the TKCX gate will evolve into the TCX gate.
\\
4)Ternary \emph{K}-Controlled-NOT (TKCNOT) gate

A TKCNOT gate ${C_{{i_1},{i_2}, \cdot  \cdot  \cdot ,{i_k};j}}({A_l})$  is defined as follows
\begin{equation}\label{e8}
\begin{array}{l}
 \bullet \;if\;j \ne l,then\;{P_l} = {C_{{i_1},{i_2}, \cdot  \cdot  \cdot ,{i_k};j}}({A_l}) = {A_l}.\\
 \bullet \;if\;j = l,and\;{A_{{i_1}}} = {A_{{i_2}}} =  \cdot  \cdot  \cdot  = {A_{{i_k}}} = 2,then\;{P_j} = {A_j}{ \oplus _3}1;else\;{P_j} = {A_j}.
\end{array}
\end{equation}
5)Generalized Controlled X(GCX) Gate

For a GCX($m'$) gate and a two-qutrit $\left| {m,A} \right\rangle $, When $m'$ is fixed as a special control bit and $m' = m$, ${X^{(ij)}}$ will be acted on target bit A; Otherwise GCX($m'$) gate will not affect the state $\left| {m,A} \right\rangle $,for $m',m \in \{ 0,1,2\}$.

\subsection{TQOTP scheme}

By referring to QOTP\cite{BR03}  and combining it with ternary quantum information technology, we design a TQOTP scheme. Suppose an \emph{n}-qutrit encryption operator is denoted as
\begin{equation}\label{e9}
{U_k} \in \{ {X^\alpha }{H^\beta }{Z^\delta }|\alpha ,\beta ,\delta  \in {\{ 0,1,2\} ^n}\}
\end{equation}
Where \emph{X}, \emph{H}, and \emph{Z} all represent ternary quantum gates, and ${X^\alpha } =  \otimes _{i = 1}^n{X^{\alpha (i)}}$, ${H^\beta } =  \otimes _{i = 1}^n{H^{\beta (i)}}$, ${Z^\delta } =  \otimes _{i = 1}^n{Z^{\delta (i)}}$. Obviously, we can use 3\emph{n} random natural numbers and patulous bit-wise \emph{XHZ} gates to encrypt an \emph{n}-qutrit quantum state. The encryption operator ${U_k}$ satisfies a uniform distribution of ${3^{3n}}$ unitary matrices. Thus, the probability ${p_k}$, which means the probability of choosing ${U_k}$, is defined as ${p_k} = {1 \mathord{\left/
 {\vphantom {1 {{3^{3n}}}}} \right.
 \kern-\nulldelimiterspace} {{3^{3n}}}}$. Actually ${U_k}$  is a ${3^n} \times {3^n}$ unitary matrix. The encryption procedure is $\left| {{\psi _c}} \right\rangle  = {U_k}\left| \varphi  \right\rangle  = {X^\alpha }{H^\beta }{Z^\delta }\left| {{\varphi _m}} \right\rangle $, and the decryption procedure is  $\left| {{\psi _m}} \right\rangle  = U_k^\dag \left| {{\psi _c}} \right\rangle  = U_k^\dag {X^\alpha }{H^\beta }{Z^\delta }\left| {{\varphi _m}} \right\rangle  = {Z^\delta }{H^\beta }{X^\alpha }{X^\alpha }{H^\beta }{Z^\delta }\left| {{\varphi _m}} \right\rangle  = \left| {{\varphi _m}} \right\rangle $.
\\
\textbf{Remark 1}. With respect to our TQOTP scheme, there are some improvements which we can do. These improvements include determining how to combine the TQOTP scheme with quantum key distribution, and construct the encryption operator ${U_k}$, and so on. By improving the TQOTP scheme, we can obtain an information-theoretically-secure scheme. These interesting problems will be discussed in future work.

\subsection{QHE scheme}

\textbf{Definition 2}. A QHE scheme is composed of four algorithms\cite{L13}: a key generation algorithm, an encryption algorithm, a decryption algorithm and an evaluation algorithm.

Compared with the usual quantum encryption scheme, the QHE scheme has a fourth evaluation algorithm, which is used to process the quantum ciphertext without decrypting it. The purpose of the evaluation algorithm is mainly to construct a homomorphic unitary operation, which can be acted on given quantum ciphertext, according to a given unitary operation. After a user has decrypted the result of the evaluation algorithm, he will obtain the same result of the corresponding operation in plaintext. So how to construct a homomorphism operation on given quantum ciphertext is the key issue in constructing the QHE scheme.
\\
\textbf{Remark 2}. Suppose ${\rho _c}$ and ${\sigma _m}$ are the forms of density matrices for quantum states. $T_\Delta $ is a set of permitted quantum operators. The evaluation algorithm can be described as follows: according to the key and the given operator $T \in {T_\Delta }$, it generates another quantum operator \emph{T}' and performs it on the ciphertext ${\rho _c}$. In this case, the operator \emph{T}' is related to the operator \emph{T} and the key. The operator \emph{T} can be regarded as a desired operation on the plaintext ${\sigma _m}$. The operation \emph{T}' corresponding to \emph{T} is performed on the ciphertext ${\rho _c}$, and can implement the desired operation on the plaintext ${\sigma _m}$.

\section{TQHE scheme}

\subsection{One-qutrit TQHE scheme}

1)One-qutrit rotation gates

For a set of permitted quantum operators $\left\{ {R_\partial ^{(ij)}(\theta )|\partial  \in \left\{ {y,z} \right\},ij \in \left\{ {01,02} \right\}} \right\}$, we only consider four forms, $R_y^{(01)}(\theta ),R_y^{(02)}(\theta ),R_z^{(01)}(\theta )$ and $R_z^{(02)}(\theta )$, in the first TQHE scheme. The evaluation algorithm will construct the corresponding homomorphic operator $\Re _\partial ^{(ij)}$, such that
\begin{equation}\label{e10}
\Re _\partial ^{(ij)}{X^\alpha }{H^\beta }{Z^\delta } = {X^\alpha }{H^\beta }{Z^\delta }R_\partial ^{(ij)}(\theta )
\end{equation}
It is deduced that $\Re _\partial ^{(ij)} = {X^\alpha }{H^\beta }{Z^\delta }R_\partial ^{(ij)}(\theta ){Z^\delta }{H^\beta }{X^\alpha }$. The first TQHE scheme is shown as follows.

${KeyGenAlgorithm}$:Randomly generate three key numbers $\alpha ,\beta ,\delta  \in \left\{ {0,1,2} \right\}$;

$EncryptAlgorithm:{\rho _c} = {X^\alpha }{H^\beta }{Z^\delta }{\sigma _m}{Z^\delta }{H^\beta }{X^\alpha }$;

$DecryptAlgorithm:{\sigma _m} = {Z^\delta }{H^\beta }{X^\alpha }{\rho _c}{X^\alpha }{H^\beta }{Z^\delta }$;

$EvaluateAlgorithm$:According to $\alpha ,\beta ,\delta $  and $R_\partial ^{(ij)}(\theta )$, it performs homomorphic  operator $\Re _\partial ^{(ij)} = {X^\alpha }{H^\beta }{Z^\delta }R_\partial ^{(ij)}(\theta ){Z^\delta }{H^\beta }{X^\alpha }$  on the given ciphertext ${\rho _c}$ without a decryption process. The output state of $EvaluateAlgorithm$ is
\begin{equation}\label{e11}
\Re _\partial ^{(ij)}{\rho _c}{(\Re _\partial ^{(ij)})^\dag } = {X^\alpha }{H^\beta }{Z^\delta }(R_\partial ^{(ij)}(\theta ){\sigma _m}{(R_\partial ^{(ij)}(\theta ))^\dag }){Z^\delta }{H^\beta }{X^\alpha }
\end{equation}
For Eq.\ref{e11}, a brief verification is given as follows,
\begin{equation}\label{e12}
\begin{array}{l}
\Re _\partial ^{(ij)}{\rho _c}(\Re {_\partial ^{(ij)})^\dag} = {X^\alpha }{H^\beta }{Z^\delta }R_\partial ^{(ij)}(\theta ){Z^\delta }{H^\beta }{X^\alpha }{X^\alpha }{H^\beta }{Z^\delta }{\sigma _m}{Z^\delta }{H^\beta }{X^\alpha }(\Re {_\partial ^{(ij)})^\dag}\\
\;\;\;\;\;\;\;\;\;\;\;\;\;\; = {X^\alpha }{H^\beta }{Z^\delta }R_\partial ^{(ij)}(\theta ){\sigma _m}{Z^\delta }{H^\beta }{X^\alpha }(\Re {_\partial ^{(ij)})^\dag}\\
\;\;\;\;\;\;\;\;\;\;\;\;\;\; = {X^\alpha }{H^\beta }{Z^\delta }R_\partial ^{(ij)}(\theta ){\sigma _m}{Z^\delta }{H^\beta }{X^\alpha }{X^\alpha }{H^\beta }{Z^\delta }(R_\partial ^{(ij)}(\theta ))^\dag {Z^\delta }{H^\beta }{X^\alpha }\\
\;\;\;\;\;\;\;\;\;\;\;\;\;\; = {X^\alpha }{H^\beta }{Z^\delta }(R_\partial ^{(ij)}(\theta ){\sigma _m}{(R_\partial ^{(ij)}{(\theta )})^\dag }{Z^\delta }{H^\beta }{X^\alpha }
\end{array}
\end{equation}
So Eq.\ref{e11} holds. After the user decrypts the state of Eq.\ref{e11}, he will obtain the desired-state as follows
\begin{equation}\label{e13}
  \begin{array}{l}
{Z^\delta }{H^\beta }{X^\alpha }\Re _\partial ^{(ij)}{\rho _c}(\Re {_\partial ^{(ij)})^\dag}{X^\alpha }{H^\beta }{Z^\delta }\\
\;\;\;\;\;\;\;\;\;\;\;\;\;\;\;\;\;\;\; = {Z^\delta }{H^\beta }{X^\alpha }{X^\alpha }{H^\beta }{Z^\delta }(R_\partial ^{(ij)}(\theta ){\sigma _m}(R_\partial ^{(ij)}{(\theta ))^\dag }){Z^\delta }{H^\beta }{X^\alpha }{X^\alpha }{H^\beta }{Z^\delta }\\
\;\;\;\;\;\;\;\;\;\;\;\;\;\;\;\;\;\;\; = R_\partial ^{(ij)}(\theta ){\sigma _m}(R_\partial ^{(ij)}{(\theta ))^\dag }
\end{array}
\end{equation}
Obviously, according to Eq.\ref{e13} the output state of $EvaluateAlgorithm$ is just the result of the  operator $R_\partial ^{(ij)}(\theta )$ acting on the plaintext ${\sigma _m}$ after the user decrypts it.

For a instance, suppose the user plaintext state is $\left| 0 \right\rangle  = {\left( {1,0,0} \right)^T}$, the user-desired operator on the plaintext is $R_y^{(01)}(\pi ) = \left( {\begin{array}{*{20}{c}}
0&0&{ - 1}\\
0&1&0\\
1&0&0
\end{array}} \right)$, and the encryption operator ${U_k}$ is ${X^2}{H^0}{Z^1} = \frac{1}{{\sqrt 2 }}\left( {\left. {\begin{array}{*{20}{c}}
1&{ - 1}&0\\
0&0&{\sqrt 2 }\\
1&1&0
\end{array}} \right)} \right.$. Thus the homomorphic operator $\Re _y^{(01)}$ is $\frac{1}{2}\left( {\left. {\begin{array}{*{20}{c}}
1&{ - \sqrt 2 }&{ - 1}\\
{\sqrt 2 }&0&{\sqrt 2 }\\
{ - 1}&{ - \sqrt 2 }&1
\end{array}} \right)} \right.$. The output state of $EvaluateAlgorithm$ is ${Z^1}{H^0}{X^2}\Re _y^{(01)}{X^2}{H^0}{Z^1}\left| 0 \right\rangle  = {\left( {0,0,1} \right)^T} = R_y^{(01)}(\pi )\left| 0 \right\rangle $, which is exactly the output of the user-desired operator $R_y^{(01)}(\pi )$ acting on the plaintext $\left| 0 \right\rangle $. Moreover, no decryption is performed during the computing of $EvaluateAlgorithm$. Thus, the scheme satisfies the definition in section 2.4, and the result of the homomorphism operator is precisely the user-desired output state.
\\
2)General one-qutrit gates

An arbitrary $3 \times 3$ unitary matrix U can be decomposed into some $R_\partial ^{(ij)}$ terms (see Eq.\ref{e7}). Accordance to Eq.\ref{e10}, we can obtain the homomorphism operator $\Re _\partial ^{(ij)}$ corresponding to $R_\partial ^{(ij)}$ acting on the quantum plaintext state. Thus the homomorphic operator ${\Re _U}$  corresponding to the unitary matrix U can synthesized by eight $\Re _\partial ^{(ij)}$. Referring to the first TQHE scheme, we can propose the second TQHE scheme for an arbitrary $3 \times 3$ unitary matrix U. It can be seen that the operator ${\Re _U}$ is a little complicated. As a result it will influence the execution efficiency of $EvaluateAlgorithm$.

\subsection{TQHE scheme for GCX gate}

\begin{figure}
  \centering
  \includegraphics[width=6.76cm,height=1.50cm]{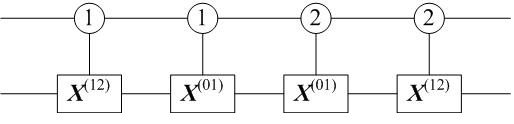}
  \caption{Synthesis of TSUM gate.}\label{f1}
\end{figure}

\begin{figure}
  \centering
  \includegraphics[width=9.35cm,height=1.53cm]{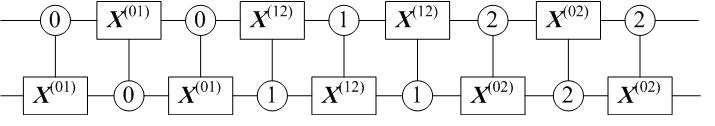}
  \caption{Synthesis of TSWAP gate.}\label{f2}
\end{figure}

\begin{figure}
  \centering
  \includegraphics[width=5.27cm,height=1.5cm]{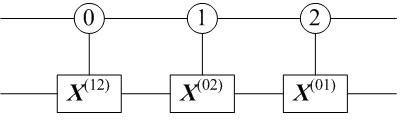}
  \caption{Synthesis of TXOR gate.}\label{f3}
\end{figure}

The Cartan Decomposition of one-qutrit gate is not unique, so the choice of one-qutrit elementary gates is not unique either. Refs.23 and 24 regard GCX gate and extended rotation gate as a two-qutrit elementary gate, respectively. In this paper, we choose the GCX gate as a ternary quantum elementary gate, which is universal for \emph{n}-qutrit quantum computing, when it is assisted by arbitrary one-qutrit gates. The common two-qutrit gates: TCX, TSWAP, TSUM (TFeynman), TXOR, and TShift gate can be synthesized by some GCX gates without auxiliary one-qutrit gates. For example, the TSUM (TFeynman) gate is synthesized by four GCX gates, as shown in Fig.\ref{f1}. Likewise, the TSWAP gate is synthesized by nine GCX gates (see Fig.\ref{f2}) and the TXOR gate synthesized by three GCX gates (see Fig.\ref{f3}).

When the control bit $m'$ is set to $\left|0\right\rangle$ state (or $\left|1\right\rangle$ or $\left|2\right\rangle$ ) and $m' = m$, the effect of the GCX(\emph{m}' ) gate is that the operator ${X^{(ij)}}$ will be applied to the target bit \emph{A}. Namely,
\begin{equation}\label{e14}
GCX(\emph{m}')\left| {m,A} \right\rangle  = ({I_3} \otimes {X^{(ij)}})\left| {m,A} \right\rangle  = \left| m \right\rangle  \otimes {X^{(ij)}}\left| A \right\rangle \
\end{equation}
If $m' \ne m$, then the operator ${X^{(ij)}}$ will be equivalent to ${I_3}$ without affecting the target bit A. Therefore the homomorphism operator of the GCX(\emph{m}') gate is exactly the similar with the operators ${X^{(ij)}} \in \left\{ {{X^{(01)}},{X^{(02)}},{X^{(12)}}} \right\}$ (equivalent to ${X^{(ij)}} \in \left\{ {{X^{(0)}},{X^{(1)}},{X^{(2)}}} \right\}$). In order to better describe the performance of the GCX(\emph{m}') gate acting on the two-qutrit $\left| {m,A} \right\rangle $, the function of $f(m',m)$ is defined as follows
\begin{equation}\label{e15}
  \tau  \equiv f(m',m) \equiv \left\{ {\begin{array}{*{20}{c}}
{{I_3},\quad \quad m' = m}\\
{{{{X^{(ij)}}}^\dag },m' \ne m}
\end{array}} \right.
\end{equation}
So the third TQHE scheme associated with the GCX gate is described as follows.

$KeyGenAlgorithm$:Randomly generate three numbers $\alpha ,\beta ,\delta  \in {\left\{ {0,1,2} \right\}^2}$;

$EncryptAlgorithm:{\rho _c} = {X^\alpha }{H^\beta }{Z^\delta }{\sigma _m}{Z^\delta }{H^\beta }{X^\alpha }$;

$DecryptAlgorithm:{\sigma _m} = {Z^\delta }{H^\beta }{X^\alpha }{\rho _c}{X^\alpha }{H^\beta }{Z^\delta }$;

$EvaluateAlgorithm$:According to $\alpha ,\beta ,\delta ,i$ and $j$, it computes the homomorphic operator ${X_{(ij)}} = {X^\alpha }{H^\beta }{Z^\delta }({I_3} \otimes {X^{(ij)}}\tau ){Z^\delta }{H^\beta }{X^\alpha }$. Then the corresponding operator ${X_{(ij)}}$ will be acted on the given ciphertext ${\rho _c}$ without performing $DecryptAlgorithm$. The output state of $EvaluateAlgorithm$ is
\begin{equation}\label{e16}
  {X_{(ij)}}{\rho _c}X_{(ij)}^\dag  = {X^\alpha }{H^\beta }{Z^\delta }(({I_3} \otimes {X^{(ij)}}\tau ){\sigma _m}{({I_3} \otimes {X^{(ij)}}\tau )^\dag }){Z^\delta }{H^\beta }{X^\alpha }
\end{equation}
Obviously it is just the same as the operator ${I_3} \otimes {X^{(ij)}}\tau $ acting on the plaintext ${\sigma _m}$. The scheme based on the GCX gate satisfies the QHE scheme’ demand.

For example, suppose that the user's plaintext state is $\left|{02}\right\rangle  = {\left( {0,0,1,0,0,0,0,0,0} \right)^T}$. For the user-desired GCX(\emph{m}'), the control bit \emph{m}' is set to 0 and operator \emph{X} is specified as ${X^{(02)}}\tau $ (equivalent to ${I_3} \otimes {X^{(02)}},\tau  = f(m',m) = {I_3}$). The encryption operator is denoted as ${\varepsilon _{{3^2} \times {3^2}}} = {X^{12}}{H^{02}}{Z^{01}} = ({X^1} \otimes {X^2})({H^0} \otimes {H^2})({Z^0} \otimes {Z^1})$, that is:${\varepsilon _{{3^2} \times {3^2}}} = \frac{1}{2}\left( {\left. {\begin{array}{*{20}{c}}
0&0&0&0&0&0&0&0&{0.5}\\
0&0&0&0&0&0&{ - \sqrt 2 }&{\sqrt 2 }&0\\
0&0&0&0&0&0&{\sqrt 2 }&{ - \sqrt 2 }&0\\
0&0&{ - \sqrt 2 }&0&0&{ - \sqrt 2 }&0&0&0\\
1&{ - 1}&0&1&{ - 1}&0&0&0&0\\
{ - 1}&{ - 1}&0&{ - 1}&{ - 1}&0&0&0&0\\
0&0&{ - \sqrt 2 }&0&0&{\sqrt 2 }&0&0&0\\
1&{ - 1}&0&{ - 1}&1&0&0&0&0\\
{ - 1}&{ - 1}&0&1&1&0&0&0&0
\end{array}} \right)} \right.$.
Thus, the homomorphic operator ${X_{(ij)}}$ is ${I_3} \otimes u\tau $, where $u = \frac{1}{2}\left( {\begin{array}{*{20}{c}}
0&{ - \sqrt 2 }&{\sqrt 2 }\\
{ - \sqrt 2 }&1&1\\
{\sqrt 2 }&1&1
\end{array}} \right)$ and $\tau  = {I_3}$. The output state is
$
 {Z^{01}}{H^{02}}{X^{12}}{X_{(ij)}}{X^{12}}{H^{02}}{Z^{01}}\left| {02} \right\rangle  = {\left( {1,0,0,0,0,0,0,0,0} \right)^T} = \left| {00} \right\rangle  = ({I_3} \otimes {X^{(02)}}\tau )\left| {02} \right\rangle
$
After a user has decrypted the result of $EvaluateAlgorithm$, he will obtain the same result  as the user-desired operator ${X^{(02)}}\tau $ acting on the plaintext ${\sigma _m}$. The scheme is in accord with the definition in section 2.4.

\subsection{\emph{n}-qutrit TQHE scheme}

Any \emph{n}-qutrit quantum gate can be expressed as a ${3^n} \times {3^n}$ unitary matrix. Based on one/two-qutrit quantum gates, there are plenty of research opportunities for \emph{n}-qutrit quantum gates by using permutation group theory and Cosine-Sine Decomposition (CSD). According to permutation group theory, all \emph{n}-qutrit($(n \ge 2)$) quantum circuits can be generated by a group of two-qutrit gates: SWAP, NOT and 1-controlled-NOT gates without ancillary qutrits\cite{YXSP06,YHSP10}. Obviously, it is only a construction-based algorithm. In terms of CSD, a ${3^n} \times {3^n}$ unitary matrix can be synthesized by 12 controlled-\emph{U} gates, 12 Dual-Shift gates, $3^n$ (\emph{N}-1)-controlled rotation gates and $2 \cdot (n - 1) \cdot {3^{n - 1}}$  TX gates\cite{KP06}. The (\emph{N}-1)-controlled rotation gate is defined as follows
\begin{equation}\label{e18}
  (N - 1)CR({A_1}, \cdot  \cdot  \cdot {A_n};B) = \left\{ {\begin{array}{*{20}{c}}
{({A_1}, \cdot  \cdot  \cdot ,{A_{n - 1}};RB),{A_i} = 2,1 \le i \le n - 1}\\
{({A_1}, \cdot  \cdot  \cdot ,{A_{n - 1}};B),\quad \quad\quad\quad \quad \quad \;\;\;others}
\end{array}} \right.
\end{equation}
For ${A_i},B \in \left\{ {0,1,2} \right\}$, $R \in \left( {\begin{array}{*{20}{c}}
{\cos \theta }&{ - \sin \theta }&0\\
{\sin \theta }&{\cos \theta }&0\\
0&0&1
\end{array}} \right)$ or $\left( {\begin{array}{*{20}{c}}
1&0&0\\
0&{\cos \theta }&{ - \sin \theta }\\
0&{\sin \theta }&{\cos \theta }
\end{array}} \right)$. No matter which method is used to synthesize a ${3^n} \times {3^n}$ unitary matrix, there always exists extremely complex process, a giant number gates and very low efficiency. At present, this is a difficulty and hot issue of synthesis of multivalue quantum gate.

\begin{figure}
  \centering
  \includegraphics[width=4.17cm,height=1.55cm]{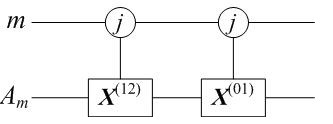}\\
  \caption{Synthesis of TNOT gate ${C_j}({A_m})$.}\label{f4}
\end{figure}

\begin{figure}
  \centering
  \includegraphics[width=4.16cm,height=2.39cm]{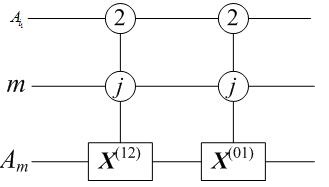}\\
  \caption{Synthesis of Ternary 1-controlled NOT gate ${C_{{i_1};j}}({A_m})$.}\label{f5}
\end{figure}

According to Ref.25 and section 3.2, we describe a process of how to construct an \emph{n}-qutrit TQHE scheme. TNOT gate and Ternary 1-controlled gate are in fact two single-shift gates with conditions (see \emph{Table} \ref{t1}). The difference is the latter requires the control bit to be 2 on base of the former’s condition (see Figs.\ref{f4} and \ref{f5}). Obviously, a TNOT gate can be synthesized by two GCX gates (see Fig.\ref{f4}). And a ternary 1-comtrolled gate can be synthesized by two conditional GCX gates (see Fig.\ref{f5}).  Therefore, we can conclude that any \emph{n}-qutrit ($n \ge 2$) logic circuits can be synthesized by some GCX gates.

Suppose the user will perform the quantum circuit ${C_{{3^n} \times {3^n}}}$ on the given quantum plaintext. In principle we can get the homomorphism operator ${C_{{3^n} \times {3^n}}^{'}}$, which is synthesized by a large number of  GCX gates' homomorphic operators, such that
\begin{equation}\label{e19}
C_{{3^n} \times {3^n}}^{'}{X^\alpha }{H^\beta }{Z^\delta } = {X^\alpha }{H^\beta }{Z^\delta }{C_{{3^n} \times {3^n}}}
\end{equation}
Where $\alpha ,\beta ,\delta  \in {\left\{ {0,1,2} \right\}^n}$. So far we can present \emph{n}-qutrit QHE scheme in theory. However, the form of the operator $C_{{3^n} \times {3^n}}$ synthesized by a great deal of GCX gates is so complex that performing $EvaluateAlgorithm$ is very bad in execution efficiency and energy consumption. Thus the \emph{n}-qutrit QHE scheme does need many improvements in multivalue quantum circuit synthesis and homomorphic ${EvaluateAlgorithm}$ design.

\section{Analysis}

\subsection{Circuit synthesis}
Currently, the synthesis of multivalue quantum logic circuit mainly focuses on the simplest ternary quantum logic circuits. Although there are some research achievements, they are not very mature themselves. How to choose universal ternary quantum gate has not been well solved so far. And there are no uniform criteria for the performance analysis, complexity, cost, energy consumption, and so on, after the ternary quantum gates being synthesized by a number of GCX gates. In short, it is feasible to construct an \emph{n}-qutrit QHE scheme in theorem, but it is too difficult to construct in practice. Therefore, this paper does not actually construct such a scheme and we only describe a general construction process by the synthesis method.

In view of the current state of research on the synthesis of an \emph{n}-qutrit quantum gate (actually a ${3^n} \times {3^n}$ unitary matrix), synthesis methods of quantum circuit require further optimization and refinement. Based on quantum circuit synthesis, construction of an \emph{n}-qutrit QHE scheme is premature. The key reasons are very low performance efficiency and difficulties in constructing the homomorphic operator while executing $EvaluateAlgorithm$. Therefore, the TQHE scheme presented in the paper is limited to one-qutrit rotation gates, one-qutrit quantum gates and two-qutrit gates synthesized by some GCX gates. Theoretically, we can generalize the TQHE scheme to the \emph{n}-qutrit ($n \ge 2$) quantum gates case.

\subsection{Security}
Our TQHE schemes, which are based on the TQOTP scheme, all assume that the delegation party is honesty. There are two aspects of this representation. One is that it will not leak or snoop users’ private data in the process of performing $EvaluateAlgorithm$  in the case of knowing the decryption key. The other is that it will right execute the commissioned unitary operators without decryption. The delegation party must know the user-given operators and the secret key. Otherwise he cannot correctly construct the homomorphic operator. It should be noted that the $EvaluateAlgorithm$  is dependent on the secret key. Thus, the delegation party can decrypt it and obtain the original qutrit. This will restrict the application of these schemes. Meanwhile these schemes cannot be used in blind quantum computation. However, we can delegate the computation to the trusted party by using the TQHE schemes, which can prevent the malicious parties from obtaining the data and the result of the computation. Of course, these schemes can be also used in secure multiparty quantum computation.

In addition to the legal user and trusted delegation party, there is no party (including the eavesdropper, etc.) that is able to get the complete user data. According to our TQOTP scheme, which is used to encrypt the user’s original data, the eavesdropper correctly guesses the secret key with the probability ${p_k} = {1 \mathord{\left/
 {\vphantom {1 {{3^{3n}}}}} \right.
 \kern-\nulldelimiterspace} {{3^{3n}}}}$. In the case of one-qutrit ($n=1$), the probability ${p_k}$ is ${{{1 \mathord{\left/
 {\vphantom {1 {{3^{3n}}}}} \right.
 \kern-\nulldelimiterspace} {{3^{3n}}}} = 1} \mathord{\left/
 {\vphantom {{{1 \mathord{\left/
 {\vphantom {1 {{3^{3n}}}}} \right.
 \kern-\nulldelimiterspace} {{3^{3n}}}} = 1} {{3^3}}}} \right.
 \kern-\nulldelimiterspace} {{3^3}}} = {1 \mathord{\left/
 {\vphantom {1 {27}}} \right.
 \kern-\nulldelimiterspace} {27}} \approx 3.7\% $ . Similarly, in the case of  two-qutrit ($n=2$), the probability ${p_k}$ is ${1 \mathord{\left/
 {\vphantom {1 {{3^6}}}} \right.
 \kern-\nulldelimiterspace} {{3^6}}} \approx 0.14\% $ and, with three-qutrit ($n=3$), the probability ${p_k}$ is ${1 \mathord{\left/
 {\vphantom {1 {{3^9}}}} \right.
 \kern-\nulldelimiterspace} {{3^9}}} \approx 5.08 \times {10^{ - 3}}\% $. With the increase of original data’s length, the secret key’s length will be longer. So the probability ${p_k}$ will become smaller and smaller, and eventually it will tend to zero. It is extremely difficult for the eavesdropper to correctly guess the secret key. Additionally, the quantum key distribution protocols, such as the BB84 protocol, can be used to generate and transmit the secret key between user and delegation party, which must make our secret key unconditional secure. As a result, our schemes will be more secure.

The eavesdropper cannot obtain effective and complete information for the user encrypted data and the output state of the $EvaluateAlgorithm$. Suppose forcing to measure the intercepted data, the eavesdropper will obtain the random information about these encrypted data. Due to the quantum \emph{No-Cloning Theorem} and \emph{Heisenberg Uncertainty Principle}, along with the TQOTP scheme, the eavesdropper knows nothing about the mutual information of the unknown cipher states, namely $I({\rho _c}:Eve) \approx 0$. There is only one round of information exchange between the user and the delegation party in our TQHE schemes, rather than several rounds of information exchange in blind quantum computation. The benefits of our schemes are the high level of user’s data privacy and the reduction of the times for which sensitive data is exposed. By performing $EvaluateAlgorithm$, the delegation party will execute the corresponding homomorphic operator in the user-given cipher states. The homomorphic operator is equivalent to a new encryption operator, which acts on the user-given ciphertext. The homomorphic operator is only known by the delegation party, and the eavesdropper knows nothing about it. Without knowing the secret key and homomorphic operator, the eavesdropper does not know the secondary encrypted quantum states, i.e.,$I({\rho _c}:Eve) \approx 0$.

The binary QHE schemes presented in Ref.13 are efficient and perfectly secure, i.e., $I({\rho _c}:{\sigma _m}) = H({\rho _c}) - H({\rho _c}|{\sigma _m}) = 0$. It shows that the ciphertext is independent of the plaintext. However, a deterministic QHFE scheme necessarily incurs exponential overhead if perfect security is required\cite{YPF14}. This is very difficult to implement in practice. The TQHE schemes in this paper are not perfectly secure because of the encryption operator ${U_k}$, which is not a complete orthogonal basis in \emph{n}-qutrit Hilbert Space. As a result, both of the output quantum states of the $EvaluateAlgorithm$ and $EncryptionAlgorithm$ are not the totally mixed states. According to definition 1 in Ref.11, our TQHE schemes are all $\varepsilon $-security.

\section{Conclusion}

At present, there exist few researches in QHE. According to present research, it is the very first time to present these TQHE schemes based on TQOTP in this paper. First, we proposed a TQOTP protocol and these TQHE schemes are based on it. Second, in allusion to the ternary quantum rotation gate $R_\partial ^{(ij)}$, we had constructed the homomorphic operator $\Re _\partial ^{(ij)}$ and the first TQHE scheme. Then according to general one-qutrit gate synthesized by eight $\Re _\partial ^{(ij)}$ gates, we presented the second TQHE scheme. Third, on the basis of Ref.23 and the GCX(\emph{m}') gate which can be as a two-qutrit universal gate, we constructed the third TQHE scheme for the GCX(\emph{m}') gate. Referring to Refs.25-26 and the third TQHE scheme, we generalized to \emph{n}-qutrit case and theoretically presented the fourth TQHE scheme. Finally, we discussed two components of these schemes’ security. One is that there exists an extremely low probability that the attacker correctly guesses the secret key. Another is that the attacker knows almost nothing about two kinds of cipher quantum states, i.e., the output encrypted quantum states of the $EvaluateAlgorithm$ and $EncryptionAlgorithm$. Meanwhile, future research ideas are to find a TQOTP scheme with perfect security and to construct an asymmetric TQHE scheme, where the $EvaluateAlgorithm$  only depends on the public key but not the private key. The latter is an open problem. Maybe one can consider how to modify the quantum public-key encryption scheme in Refs.29-30, such that it becomes an asymmetric QHE scheme. If this goal were achieved, the computing on the quantum cipher state could be securely outsourced, and then blind quantum computation would be implemented in this way.

\section*{Acknowledgments}
We are grateful to Prof. Yang for warm and useful discussions and suggestions. We acknowledge supports from the project of National Natural Science Foundation of China (Grant No.61272175) and the research project of Minnan Normal University (Grant No.sk09002).

\end{document}